\documentclass[twocolumn,showpacs,preprintnumbers,amsmath,amssymb]{revtex4}
\usepackage{graphicx}% Include figure files
\usepackage{dcolumn}% Align table columns on decimal point
\usepackage{bm}% bold math

\begin{document}

\preprint{Accepted for publication in Physical Review Letters}

\title{Constraining the Spectrum of Supernova Neutrinos \\
from $\nu$-Process Induced Light Element Synthesis}% Force line breaks with \\

\author{Takashi Yoshida$^1$}\email{tyoshida@astr.tohoku.ac.jp}
\author{Toshitaka Kajino$^{2,3}$}%
% \email{kajino@nao.ac.jp}
\author{Dieter H. Hartmann$^{4}$}
\affiliation{%
$^1$Astronomical Institute, Graduate School of Science, Tohoku University,
Sendai 980-8578, Japan \\
$^2$National Astronomical Observatory of Japan, and The Graduate University
for Advanced Studies, Tokyo 181-8588, Japan \\
$^3$Department of Astronomy, Graduate School of Science, University of Tokyo,
Tokyo 113-0033, Japan \\
$^4$ Department of Physics and Astronomy, Clemson University, Clemson,
South Carolina 29634, USA
}%

%\affiliation{

\date{\today}% It is always \today, today,

\begin{abstract}
We constrain energy spectra of supernova neutrinos through the avoidance of an 
overproduction of the $^{11}$B abundance during Galactic chemical evolution.
In supernova nucleosynthesis calculations with a parametrized neutrino
spectrum as a function of temperature of $\nu_{\mu,\tau}$ and 
$\bar{\nu}_{\mu,\tau}$ and total neutrino energy, we find a strong neutrino
temperature dependence of the $^{11}$B yield.
When the yield is combined with observed abundances, the acceptable range of 
the $\nu_{\mu,\tau}$ and $\bar{\nu}_{\mu,\tau}$ temperature is 
found to be 4.8 to 6.6 MeV. 
Nonzero neutrino chemical potentials would reduce this temperature range 
by about 10\% for a degeneracy 
parameter $\eta_\nu=\mu_\nu/kT_\nu$ smaller than 3.
\end{abstract}

\pacs{26.30.+k, 97.60.Bw, 25.30.Pt}% PACS, the Physics and Astronomy
\maketitle

The light elements (Li-Be-B) are continuously produced by supernovae (SNe) 
\cite{wh90,ww95,ra02}, as well as interactions of Galactic cosmic rays (GCRs) 
with the interstellar medium (ISM), nucleosynthesis in asymptotic giant branch
(AGB) stars, and novae during Galactic chemical evolution (GCE, i.e., the 
evolution in chemical composition of stars and interstellar gas during 
Galactic history) \cite{fo00,rl00}.
In the case of boron, cosmic ray induced spallation in the ISM and supernova
ejecta dominate the production; $^{11}$B is contributed through both channels, 
while $^{10}$B production is probably exclusively due to GCRs.
The contribution from supernovae to the production of $^{11}$B can be 
calibrated with the isotopic ratio N($^{11}$B)/N($^{10}$B), 
measured with great precision in primitive meteorites.

The SN $\nu$-process plays an important
role for $^{11}$B and $^7$Li production \cite{wh90}.
The interaction of neutrinos, emitted in copious amounts during core collapse 
and the subsequent cooling phase of proto-neutron stars, with matter in the 
ejecta of SNe, contributes uniquely to GCE. 
Recent studies based on the theoretical yields derived by Woosley and 
Weaver (WW95) \cite{ww95} suggest 
that the SN contribution to the $^{11}$B abundance is significantly larger 
than that required to explain the boron evolution in the Galactic 
disk and the meteoritic $^{11}$B/$^{10}$B ratio \cite{fo00,rl00,al02}.

To match the abundance of $^{11}$B established during GCE, we previously 
assumed neutrino energy spectra to resemble Fermi-Dirac (FD) distributions 
with zero chemical potentials $\mu_\nu=0$ \cite{wh90,ww95,ra02}
and fixed neutrino temperatures of 6.0, 3.2, and 5.0 MeV for 
$\nu_{\mu,\tau}$ ($\bar{\nu}_{\mu,\tau}$), 
$\nu_{\rm e}$, and $\bar{\nu}_{\rm e}$, respectively \cite{yt04}.
The $\nu_{\mu,\tau}$ temperature of 6.0 MeV is significantly smaller than the 
8.0 MeV used in the other previous studies of the $\nu$-process 
\cite{wh90,ww95,ye00}.
This reduction is derived from an investigation of the dependence of 
the $^{11}$B yield on the total neutrino energy $E_\nu$ and the decay time 
scale $\tau_\nu$ of the neutrino flux. 
The yield is roughly proportional to $E_\nu$ and rather insensitive 
to $\tau_\nu$.
The temperature dependence was not investigated very well.

Studies of supernova explosions with detailed neutrino transport 
(e.g., \cite{jb01,kr03} and references therein) have indicated 
that emerging neutrino spectra do not closely follow FD distributions 
with $\mu_\nu=0$. Since the high-energy tail of the energy distribution 
is predominantly important for the $\nu$-process (e.g. \cite{wh90}), 
the use of FD distribution with $\mu_{\nu}=0$ may be 
justified as an approximation as long as the spectrum above 
$\varepsilon_\nu \approx 10$ MeV is a good match to
the shapes obtained in detailed transport simulations \cite{kr03,mb90,hm99}. 
However, if the $^{11}$B yield depends strongly on the neutrino 
temperatures, which have not yet been clarified theoretically, the nonzero 
chemical potentials would change the resultant $^{11}$B abundance in a
different matter from what follows from FD distributions with $\mu_{\nu} = 0$.
The purpose of this Letter is to investigate the neutrino temperature 
dependence of the SN $\nu$-process in detail, and to find out how 
robustly lower neutrino temperatures may provide the means to avoid 
overproduction of the $^{11}$B abundance in GCE and meteoritic 
$^{11}$B/$^{10}$B ratio.
% even by using more realistic FD 
% distributions which have nonzero neutrino chemical potentials.

The adopted model for SN neutrinos is guided by numerical simulations
from the literature, with a few additional simplifying assumptions. 
The neutrino luminosity is assumed to be uniformly partitioned among 
the neutrino flavors, and is assumed to decrease exponentially in time with 
a time scale $\tau_\nu = 3$ s \cite{wh90}. 
The latter assumption is not critical, because the ejected masses of 
$^{11}$B and $^7$Li are insensitive to $\tau_\nu$ \cite{yt04}. 
We initially assume that the spectra indeed obey the FD form with $\mu_\nu=0$.

\begin{figure}
\includegraphics{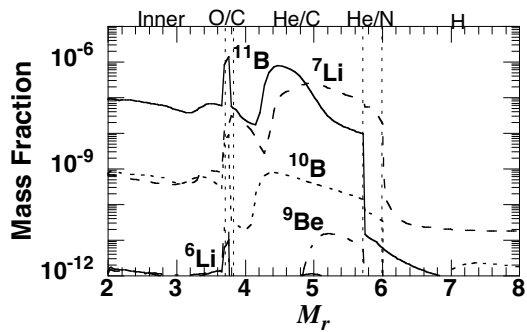}
\caption{
Mass fraction distribution of the light elements in the $16.2 M_\odot$ model
with $T_{\nu_{\mu,\tau}}=6$ MeV. The horizontal axis is the interior mass in 
units of the solar mass.
}
\end{figure}

Only the total neutrino energy $E_\nu$ and the temperatures 
$T_{\nu_{\mu,\tau}}$ are free parameters. 
The allowed range of the total neutrino energy $E_\nu$ is
\begin{equation}
1 \times 10^{53} \, {\rm ergs} \le E_\nu \le 6 \times 10^{53} \, {\rm ergs,}
\label{eq1}
\end{equation}
which includes the reduced range 
$2.4 \times 10^{53}$ ergs $ \le E_\nu \le 3.5 \times 10^{53}$ ergs, 
corresponding to the estimated range in gravitational binding energy of 
a neutron star with mass $\sim 1.4 M_\odot$ \cite{lp01}.
The considered range of the neutrino temperature $T_{\nu_{\mu,\tau}}$ is
\begin{equation}
4.0 \, {\rm MeV} \le T_{\nu_{\mu,\tau}} \le 9.0 \, {\rm MeV .}
\label{eq2}
\end{equation}
Temperatures of the $\nu_{\rm e}$ and $\bar{\nu}_{\rm e}$, 
$T_{\nu_{\rm e}}$ and $T_{\bar{\nu}_{\rm e}}$, 
are less important for the $\nu$-process of the light elements, and we
set their values to 3.2 and 5.0 MeV, respectively \cite{yt04}.

The SN model used in this work is identical to that described by \cite{yt04}.
We use progenitor model 14E1, with a mass at explosion of $16.2 M_\odot$ 
\cite{sn90}, corresponding to SN 1987A.
The propagation of a shock wave during the SN explosion is
followed with a spherically symmetric Lagrangian PPM code
\cite{cw84,sn92}.
The explosion energy and the mass cut are set to $1 \times 10^{51}$ ergs
and $1.61 M_\odot$, respectively.
Then, we calculate explosive nucleosynthesis by postprocessing
as described in \cite{yt04}. 
The reaction rates of the $\nu$-process are derived by interpolating 
the logarithmic values of the cross sections listed in the tables of 
\cite{hw92}. 
This setup determines the thermodynamic histories of the various
mass shells that ultimately constitute the supernova ejecta (no fall back), and
the $\nu$-process yields within the ejecta are then determined through the 
cross sections as soon as the time- and energy-dependent neutrino flux is 
specified.
We calculate the yields for a parameter grid with 126 points, with steps
of $1 \times 10^{53}$ ergs in $E_\nu$ and steps of 0.25 MeV 
in $T_{\nu_{\mu,\tau}}$.
Shown in Fig. 1 is an example of the produced mass fractions for
$E_\nu$ = $3 \times 10^{53}$ ergs and $T_{\nu_{\nu,\tau}}=6$ MeV.

\begin{figure}
\includegraphics{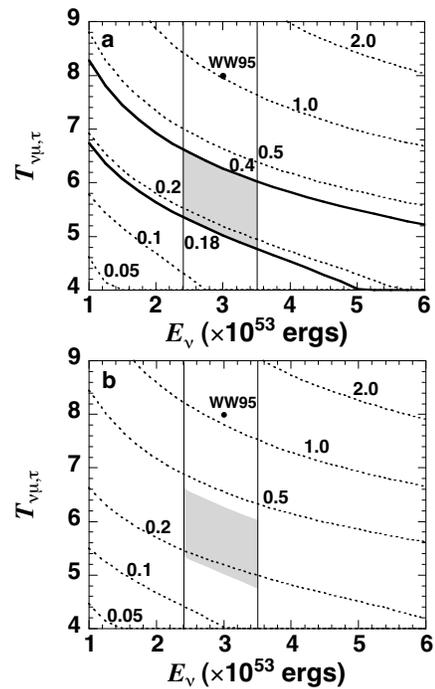}
\caption{
Contour lines of the overproduction factor $f_\nu$ for (a) $^{11}$B and 
(b) $^7$Li in the parameter plane of total neutrino energy $E_\nu$ and 
neutrino temperature $T_{\nu_{\mu,\tau}}$ (see text for details).
The region between the two solid vertical lines indicates the energy range 
relevant for a neutron star of mass $\sim 1.4 M_\odot$ \cite{lp01}. 
The point labeled WW95 indicates the specific parameter values 
used in \cite{ww95}.
In panel (a), the region between the two solid contour lines is the range of 
ejected mass appropriate for GCE of $^{11}$B.
The shaded region is the part of parameter space in which both constraints 
(GCE yield of $^{11}$B and neutron star binding energy) are simultaneously 
satisfied.
A similar box is drawn in (b) for the case of $^7$Li. 
}
\end{figure}

Ratios of the ejected masses of $^{11}$B and $^7$Li to those of WW95
\cite{ww95}, defined as the overproduction factor $f_\nu$, are shown in 
Fig. 2 as contours in the $E_\nu$ vs $T_{\nu_{\mu,\tau}}$ plane.
For the WW95 case of $T_{\nu_{\mu,\tau}}$ = 8 MeV and 
$E_\nu$ = $3 \times 10^{53}$ ergs, we find ejected $^{11}$B and $^7$Li 
masses of $1.92 \times 10^{-6} M_{\odot}$ and 
$7.37 \times 10^{-7} M_\odot$, which are very close to the yields of 
$1.85 \times 10^{-6} M_{\odot}$ and $6.67 \times 10^{-7} M_\odot$ 
obtained with the S20A model of \cite{ww95}, respectively.
The $^{11}$B mass ratio changes between 0.038 (lower left corner of Fig 2a) 
and 2.9 (upper right corner) in the assumed ranges of $E_\nu$ 
[Eq. (\ref{eq1})] and $T_{\nu_{\mu,\tau}}$ [Eq. (\ref{eq2})]. 
The mass ratio of $^7$Li changes between 0.039
[lower left corner of Fig 2(b)] and 3.3 (upper right corner). 
Note that dependence on the explosion energy and the mass cut is weak.

We constrain the neutrino temperature $T_{\nu_{\mu,\tau}}$ by requiring
that overproduction of $^{11}$B in GCE must be avoided. 
The overproduction factor depends on details of the GCE model, 
%(such as the distribution function of stellar masses at birth), 
and ranges between 0.18 \cite{rl00} and 0.40 \cite{fo00}. 
These values are obtained by combining the solar $^{11}$B/$^{10}$B ratio
with a measure of the relative cosmic-ray and supernova contribution to
solar $^{11}$B. They are shown in Fig. 2(a) as two solid lines.
If we adopt the $1.4 M_\odot$ neutron star energy range mentioned above 
($2.4 \times 10^{53}$ ergs $\le E_\nu \le 3.5 \times 10^{53}$ ergs 
\cite{lp01}), we obtain the shaded region shown in Fig. 2(a), which implies 
that the neutrino temperature $T_{\nu_{\mu,\tau}}$ satisfies
\begin{equation}
4.8 \, {\rm MeV} \le T_{\nu_{\mu,\tau}} \le 6.6 \, {\rm MeV.}
\label{eqn4}
\end{equation}

With the neutrino temperature and total energy constrained by GCE of 
$^{11}$B, we can derive a corresponding constraint on the $^7$Li yield.
Figure 2(b) shows the shaded region corresponding to the 
$E_\nu$-$T_{\nu_{\mu,\tau}}$ limits of the shaded box in Fig. 2(a). 
This region implies an ejected mass ratio of $^7$Li between 0.19 and 0.43. 
If $^{11}$B production is indeed dominated by the contributions from
the $\nu$-process, the analysis presented above implies a predicted range of 
yields for $^7$Li, which in turn constrains the contribution to 
$^7$Li production from AGB stars and novae.

We note that the smallest value of our allowed range for $T_{\nu_{\mu,\tau}}$
is in fact smaller than the assumed value of $T_{\bar{\nu}_{\rm e}}= 5.0$ MeV.
Since the neutrinospheres of $\nu_{\rm e}$ and $\bar{\nu}_{\rm e}$ are 
larger than those of $\nu_{\mu,\tau}$ and $\bar{\nu}_{\mu,\tau}$ due to 
charged current interactions, the average energy of $\nu_{\rm e}$ and
$\bar{\nu}_{\rm e}$ are smaller than those of $\nu_{\mu,\tau}$ and
$\bar{\nu}_{\mu,\tau}$ (e.g., \cite{jb01}).
Thus, if $T_{\nu_{\mu,\tau}}$ is indeed smaller than 5.0 MeV, 
$T_{\nu_{\rm e}}$ and $T_{\bar{\nu}_{\rm e}}$ should also be smaller than 
5.0 MeV. 

We also note that neutrino oscillations would raise the contribution of 
electron neutrinos to the $^{11}$B and $^7$Li production.
If neutrino conversion between $\nu_{\rm e}$ and $\nu_{\mu,\tau}$ occurs
in the oxygen-rich layer [e.g., large mixing angle with $\theta_{13}$ large
(LMA-L) case in \cite{tw01,kam}], the rates of
charged current reactions such as $^4$He($\nu_{\rm e},{\rm e}^-p)^3$He and
$^{12}$C($\nu_{\rm e},{\rm e}^-p)^{11}$C increase, keeping the rates of 
neutral-current reactions unchanged.
The yields of $^{11}$B and $^7$Li would increase by this effect
and, thus, lower neutrino temperature is favorable for avoiding overproduction
of $^{11}$B.
Additional constraints derive from $r$-process nucleosynthesis 
in neutrino driven winds \cite{yt04}.

We use a specific stellar mass model of $\sim 20 M_\odot$ to demonstrate 
the sensitivity of the $^{11}$B and $^7$Li yields to $E_\nu$ and 
$T_{\nu_{\mu,\tau}}$.
This sensitivity can also be applied to supernova models with different
progenitor masses, because the dominant production processes for $^{11}$B 
and $^7$Li are the $\nu$-process and $\alpha$-capture reactions, 
which are insensitive to progenitor masses, specifically, the $^4$He and 
$^{12}$C abundances.
In the He-rich layer, $^7$Li is produced through the reaction sequences 
$^4$He($\nu,\nu'p)$$^3$H($\alpha,\gamma)$$^7$Li and 
$^4$He($\nu,\nu'n)$$^3$He($\alpha,\gamma)$$^7$Be($n,p)^7$Li.
Most of $^{11}$B is produced through $^7$Li($\alpha,\gamma)$$^{11}$B and 
$^7$Be($\alpha,\gamma)^{11}$C($\beta^{+})^{11}$B, or the $\nu$-process
$^{12}$C($\nu,\nu'p)^{11}$B in the oxygen-rich layer \cite{wh90,ww95,yt04}.
The dependence on $E_\nu$ and $T_{\nu_{\mu,\tau}}$ of the ejected masses
of $^{11}$B and $^7$Li solely relates to that of the $\nu$-process reaction
rates. The $\alpha$-capture rates do not 
depend on the neutrino parameters, and the abundances of $^4$He and $^{12}$C 
are solely determined during the precollapse stage.
Thus, the ejected masses of $^{11}$B and $^7$Li are proportional to the 
$\nu$-process reaction rates in accordance with the values of $E_\nu$ and 
$T_{\nu_{\mu,\tau}}$.
The neutrino spectrum might depend on progenitor mass, but the extent of this
effect has not yet been established.

Many studies of $\nu$-induced nucleosynthesis assume FD
distributions with $\mu_\nu=0$ \cite{wh90,ww95,ra02,ye00,yt04}. 
However, simulations of neutrino transport in supernova explosions show that 
the energy spectra are better represented by FD distributions with nonzero
chemical potential \cite{kr03,mb90,hm99}. 
Therefore, we now consider the effect of nonzero chemical potentials 
within a semianalytic model. 
We assume that the energy dependence of the neutrino-matter interaction 
cross sections can be expressed as a simple power law
$\sigma(\epsilon) = \sigma_0 \epsilon^\alpha$.
The specific case of $\alpha=2$ was discussed in \cite{hm99}. Here we extend
their discussion to a wider range of values for $\alpha$. We assume that the 
energy spectra are exact FD distributions, specified by values of 
temperature $T_\nu$ and degeneracy parameter $\eta_\nu=\mu_\nu/kT_\nu$, 
where $k$ is the Boltzmann constant. 
With the following definition of a moment function 
\begin{equation}
F_q(\eta_\nu)=\frac{1}{2 \pi^2 (\hbar c)^3}
\int_{0}^{\infty} \frac{x^q dx}{\exp(x-\eta_\nu)+1} \, ,
\label{eq5}
\end{equation}
the neutrino number density 
$n_\nu(T_\nu,\eta_\nu)$ and energy density 
$\epsilon_\nu(T_\nu,\eta_\nu)$ can be expressed as
$F_2(\eta_\nu) (kT_\nu)^3$ and $F_3(\eta_\nu) (kT_\nu)^4$, respectively.
For a neutrino spectrum specified by $T_\nu$ and $\eta_\nu$, the average cross 
section $\sigma(T_\nu,\eta_\nu)$ is then given by
$[F_{\alpha+2}(\eta_\nu)/F_2(\eta_\nu)] \sigma_0 (kT_\nu)^\alpha$,
and is related to the average cross section one would obtain from a spectrum 
with zero chemical potential $\sigma(T_\nu,0)$ as
$[F_{\alpha+2}(\eta_\nu) F_2(0)]/[F_{\alpha+2}(0)F_2(\eta_\nu)] 
\sigma(T_\nu,0)$.

\begin{figure}
\includegraphics{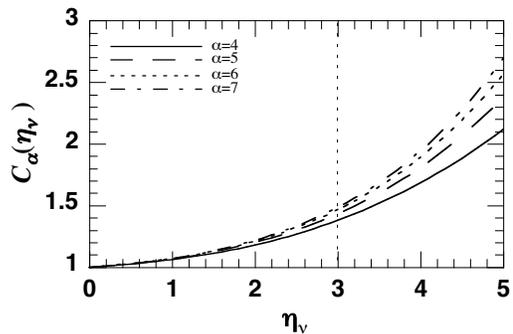}
\caption{
The scaling function with nonzero chemical potential $C_\alpha(\eta_\nu)$ 
defined by Eq. (\ref{eq9}).
The particular values of $\alpha$ shown are 4, 5, 6, and 7.
}
\end{figure}

The reaction rate for any $\nu$-process reaction under consideration at 
a given time $t$ and at a distance $r$ from the source 
is given by $\lambda(T_\nu,\eta_\nu; t)
= \sigma(T_\nu,\eta_\nu) \phi(T_\nu,\eta_\nu;t)$, 
where the neutrino number flux is 
\begin{equation}
\phi(T_\nu,\eta_\nu;t) = \frac{1}{4 \pi r^2}
\frac{E_\nu}{\frac{F_3(\eta_\nu)}{F_2(\eta_\nu)}kT_\nu}
\frac{1}{\tau_\nu} \exp \left( -\frac{t-r/c}{\tau_\nu} \right) .
\label{eq8}
\end{equation}
Note that $\phi(T_\nu,\eta_\nu;t)$ is a function of not only $T_\nu$
but also $\eta_\nu$ because the average neutrino energy depends on 
$T_\nu$ and $\eta_\nu$: 
the average energy per neutrino is $\langle\varepsilon_\nu\rangle = 
(F_3(\eta_\nu)/F_2(\eta_\nu))kT_\nu$ and
$F_3(0)/F_2(0)=3.1514$ for $\eta_\nu=0$.
The reaction rate $\lambda(T_\nu,\eta_\nu;t)$ can then be expressed as
\begin{equation} 
\lambda(T_\nu,\eta_\nu;t)=C_\alpha(\eta_\nu)\lambda(T_\nu,0;t)
\label{eq11}
\end{equation}
where $C_\alpha(\eta_\nu)$ is the scaling function
\begin{equation}
C_\alpha(\eta_\nu)=\frac{F_{\alpha+2}(\eta_\nu)}{F_{\alpha+2}(0)}
\frac{F_3(0)}{F_3(\eta_\nu)} \, ,
\label{eq9}
\end{equation}
and both $\lambda(T_\nu,\eta_\nu;t)$ and $\lambda(T_\nu,0;t)$ have the
same $T_\nu$ dependence $\propto T_\nu^{\alpha-1}$.

We now apply this semianalytic model.
First, we determine the effective power law indices for the total 
neutral-current cross sections on $^{56}$Fe and $^{58}$Fe from the 
calculations presented in \cite{tk01}. We find indices of 3.7 and 
3.8, respectively. This implies
$\sigma(T_\nu,3)/\sigma(T_\nu,0)$ =1.72(1.73) for $\alpha$=3.7(3.8),
consistent with the values reported in \cite{tk01}.
We also evaluate the power law indices $\alpha$ of the cross sections of 
$^4$He($\nu,\nu'p)^3$H and $^{12}$C($\nu,\nu'p)^{11}$B by fitting the
cross sections in \cite{hw92}.
For $^4$He($\nu,\nu'p)^3$H and $^{12}$C($\nu,\nu'p)$$^{11}$B, we find
$\alpha$ = 6.7 and 5.9, respectively.
These indices are much larger than the values obtained for reactions with
the larger nuclear systems $^{56}$Fe and $^{58}$Fe, indicating a significant
mass number dependence of $\alpha$. We therefore evaluate $C_\alpha(\eta_\nu)$ 
for $\alpha$ ranging from 4 to 7.

Figure 3 shows $C_\alpha(\eta_\nu)$ as a function of $\eta_\nu$ for various
values of $\alpha$, indicating that the rates of the $\nu$-process can vary
substantially with the adopted values for these two key parameters. 
The production of $^7$Li and $^{11}$B is proportional to the reaction retes
of $^4$He($\nu,\nu'p)^3$H and $^{12}$C($\nu,\nu'p)^{11}$B, which have similar
values of $\alpha$ (see above), so that the ejected masses of $^{11}$B and
$^{7}$Li in the case of $\eta_\nu=3$ would be increased by about 50$\%$ in 
comparison to the yield obtained for $\eta_\nu=0$.

When we allow for nonzero chemical potentials, the corresponding range of 
 $T_{\nu_{\mu,\tau}}$ derived from the GCE constraint for $^{11}$B changes.
Consider the relation between the neutrino temperatures derived from a given
yield obtained with either nonzero chemical potential $T_\nu$ or with zero 
chemical potential $T_{\nu0}$ by enforcing $\lambda(T_\nu,\eta_\nu;t)
=\lambda(T_{\nu0},0;t)$. 
The ratio of these two temperatures is given as
\begin{equation}
\frac{T_\nu}{T_{\nu0}}=C_\alpha(\eta_\nu)^{-1/(\alpha-1)} \, .
\label{eq10}
\end{equation}
%and shown in Fig. 3. 
For nonzero chemical potentials, $T_\nu/T_{\nu0}$ is a monotonically
decreasing function of $\eta_\nu$. 
In the case of $\eta_\nu=3$ we find $T_\nu/T_{\nu0} = 0.90$ for $\alpha=4$ and
0.94 for $\alpha=7$.
This implies that the neutrino temperature satisfying the GCE production 
constraint of $^{11}$B is reduced to 4.3 MeV 
$\leq T_{\nu_{\mu,\tau}}(\eta_\nu=3) \leq 5.9$ MeV, about $6\% \sim 10$\% 
smaller than the range inferred for $\eta_\nu=0$ [Eq. (\ref{eqn4})]. 
Likewise, $T_{\nu_{\rm e}}$ and $T_{\bar{\nu}_{\rm e}}$
would be reduced by a comparable fraction.
In the case of negative $\eta_\nu$, $T_\nu/T_{\nu0}$ increases very weakly,
e.g., $T_\nu/T_{\nu0}=1.015$ for $\eta_\nu=-3$.

In summary, the ejected masses of $^{11}$B and $^7$Li increase with 
$\nu_{\mu,\tau}$ and $\bar{\nu}_{\mu,\tau}$ temperature through 
the energy dependence of the cross sections of the $\nu$-process: 
%larger neutrino temperatures produce larger yields of 
%both $^{11}$B and $^7$Li.
This dependence ($\propto T_\nu^{\alpha - 1}$)is stronger than 
the dependence on the total neutrino energy.
To reproduce the supernova contribution of $^{11}$B within the framework of 
GCE, neutrino temperature is constrained to 
4.8 MeV $\le T_{\nu_{\mu,\tau}}(\eta_\nu=0) \le$ 6.6 MeV. 
Nonzero neutrino chemical potential leads to a larger light element yield.
The ejected masses of $^{11}$B and $^7$Li would be increased by about 50\%
in the case of $\eta_\nu=3$. For a given yield, the required neutrino 
temperatures are reduced correspondingly, but the change is less than 10\%.
%The neutrino temperature range constrains the $^7$Li abundance in supernovae, 
%which would constrain that in AGB stars and novae in GCE.
The inferred temperature range provides a constraint on theoretical models
of neutrino transport in supernovae and constrains their $^7$Li yields, which 
imposes constraints on contributions from AGB stars and novae to Galactic $^7$Li.

%\clearpage 

%\begin{acknowledgments}
We thank Koichi Iwamoto, Ken'ichi Nomoto, and Toshikazu 
Shigeyama for providing the data for progenitor 
model 14E1 and for helpful discussions.
T.Y. is supported by COE Research in Tohoku University (22160028).
This work has been supported in part by the Ministry of Education, Culture, 
Sports, Science and Technology, Grants-in-Aid 
%for Scientific Research (12047233, 13640313), 
for Specially Promoted Research (13002001) and by Mitsubishi Foundation.

\end{document}